\documentclass[fleqn,usenatbib]{mnras}

\usepackage{newtxtext,newtxmath}

\usepackage[T1]{fontenc}

\DeclareRobustCommand{\VAN}[3]{#2}
\let\VANthebibliography\thebibliography
\def\thebibliography{\DeclareRobustCommand{\VAN}[3]{##3}\VANthebibliography}

\usepackage{graphicx}	
\usepackage{amsmath}	

\title[Radio spectra of pulsars]{Radio spectra of pulsars fitted with the spectral distribution function of the emission from their current sheet}

\author[H. Ardavan]{
Houshang Ardavan\thanks{Email address: ardavan@ast.cam.ac.uk}\\
Institute of Astronomy, University of Cambridge,\\ 
Madingley Road, Cambridge CB3 0HA, United Kingdom}

\date{Accepted XXX. Received YYY; in original form ZZZ}

\pubyear{2023}

\begin{document}
\label{firstpage}
\pagerange{\pageref{firstpage}--\pageref{lastpage}}
\maketitle

\begin{abstract}
In their catalogue of pulsars' radio spectra, Swainston et al.\ (2022, PASA, 39, e056) distinguish between five different forms of these spectra: those that can be fitted with (i) a simple power law, (ii) a broken power law, (iii) a low-frequency turn-over, (iv) a high-frequency turn-over or (v) a double turn-over spectrum.  Here, we choose two examples from each of these categories and fit them with the spectral distribution function of the caustics that are generated by the superluminally moving current sheet in the magnetosphere of a non-aligned neutron star.  In contrast to the prevailing view that the curved features of pulsars' radio spectra arise from the absorption of the observed radiation in high-density environments, our results imply that these features are intrinsic to the emission mechanism.  We find that all observed features of pulsar spectra (including those that are normally fitted with simple or broken power laws) can be described by a single spectral distribution function and regarded as manifestations of a single emission mechanism.  From the results of an earlier analysis of the emission from a pulsar's current sheet and the values of the fit parameters for each spectrum, we also determine the physical characteristics of the central neutron star of each considered example and its magnetosphere.
\end{abstract}



\section{Introduction}
\label{sec:introduction}

Attempts at explaining the radiation from pulsars has so far been focused mainly on mechanisms of acceleration of charged particles~\citep[see, e.g.,  the references in][]{Melrose2021}: an approach spurred by the fact that, once the relevant version of this mechanism is identified, one can calculate the electric current density associated with the accelerating charged particles involved and thereby evaluate the classical expression for the retarded potential that describes the looked-for radiation.  In the present paper, however, we evaluate the retarded potential, and hence the generated radiation field, using the macroscopic distribution of electric charge-current density that is already provided by the numerical computations of the structure of a non-aligned pulsar magnetosphere~\citep[][Section~2]{Ardavan2021}.  Both the radiation field thus calculated and the electric and magnetic fields that pervade the pulsar magnetosphere are solutions of Maxwell's equations for the same charge-current distribution.  These two solutions are completely different, nevertheless,  because they satisfy different boundary conditions: the far-field boundary conditions with which the structure of the pulsar magnetosphere is computed are radically different from the corresponding boundary conditions with which the retarded solution of these equations (i.e. the solution describing the radiation from the charges and currents in the pulsar magnetosphere) is derived (see Section~3 and the last paragraph in Section~6 of~\citealt{Ardavan2021}). 

Numerical computations based on the force-free and particle-in-cell formalisms have now firmly established that the magnetosphere of a non-aligned neutron star entails a current sheet outside its light cylinder whose rotating distribution pattern moves with linear speeds exceeding the speed of light in vacuum (see~\citealt{SpitkovskyA:Oblique,Contopoulos:2012,Tchekhovskoy:etal}; and the references in~\citealt{Philippov2022}).  However, the role played by the superluminal motion of this current sheet in generating the multi-wavelength, focused pulses of radiation that we receive from neutron stars is not generally acknowledged. Given that the superluminally moving distribution pattern of this current sheet is created by the coordinated motion of aggregates of subluminally moving charged particles~\citep[see][]{GinzburgVL:vaveaa,BolotovskiiBM:Radbcm}, the motion of any of its constituent particles is too complicated to be taken into account individually.  Only the densities of charges and currents enter the Maxwell's equations, on the other hand, so that the macroscopic charge-current distribution associated with the magnetospheric current sheet takes full account of the contributions toward the radiation that arise from the complicated motions of the charged particles comprising it.

The radiation field generated by a uniformly rotating volume element of the distribution pattern of the current sheet in the magnetosphere of a non-aligned neutron star embraces a synergy between the superluminal version of the field of synchrotron radiation and the vacuum version of the field of \v{C}erenkov radiation.  Once superposed to yield the emission from the entire volume of the source, the contributions from the volume elements of this distribution pattern that approach the observation point with the speed of light and zero acceleration at the retarded time interfere constructively and form caustics in certain latitudinal directions relative to the spin axis of the neutron star.  The waves that embody these caustics are more focused the farther they are from their source: as their distance from their source increases, two nearby stationary points of their phases draw closer to each other and eventually coalesce at infinity.  By virtue of their narrow peaks in the time domain, the resulting focused pulses thus procure frequency spectra whose distributions extend from radio waves to gamma-rays~\citep[][Table~1 and Section~5.4]{Ardavan2021}.  

This paper is concerned with the radio spectra of pulsars.  Its task is to ascertain whether the spectrum of the caustics generated by a pulsar's current sheet (Section~\ref{sec:spectrum}) can account for all five categories of spectral shapes catalogued\footnote{{\url{https://all-pulsar-spectra.readthedocs.io/en/latest/}}} by~\citet{Swainston}.  To this end, it presents fits to two examples (with the largest number of known data points) of the catalogued spectra in each category (Section~\ref{sec:fits}) and relates the values of their fit parameters to the physical characteristics of the central neutron star of the corresponding pulsar and its magnetosphere (Section~\ref{sec:connection}).

\section{Radio spectrum of the caustics generated by the superluminally moving current sheet}
\label{sec:spectrum}

The frequency spectrum of the radiation that is generated as a result of the superluminal motion of the current sheet in the magnetosphere of a non-aligned neutron star was presented, in its general form, in equation~(177) of~\citet[][Section~5.3]{Ardavan2021}. 

In a case where the magnitudes of the vectors denoted by $\boldsymbol{\cal P}_l$ and $\boldsymbol{\cal Q}_l$ in equation~(177) of~\citet{Ardavan2021} are appreciably larger than those of their counterparts, ${\bar{\boldsymbol{\cal P}}}_l$ and ${\bar{\boldsymbol{\cal Q}}}_l$, and the dominant contribution towards the Poynting flux $S_\nu$ of the radiation is made by only one of the two terms corresponding to $l=1$ and $l=2$, e.g.\ $l=2$, that equation can be written as
\begin{equation}
S_\nu=\kappa_0\, k^{-2/3}\left\vert\boldsymbol{\cal P}_2\, {\rm Ai}(-k^{2/3}\sigma_{21}^2)-{\rm i}k^{-1/3}\boldsymbol{\cal Q}_2\,{\rm Ai}^\prime(-k^{2/3}\sigma_{21}^2)\right\vert^2,
\label{E1}
\end{equation}
where Ai and ${\rm Ai}^\prime$ are the Airy function and the derivative of the Airy function with respect to its argument, respectively, $k=2\pi\nu/\omega$ is the frequency $\nu$ of the radiation in units of the rotation frequency $\omega/2\pi$ of the central neutron star, and $\kappa_0$ and $\sigma_{21}$ are two positive scalars.  The coefficients of the Airy functions in the above expression stand for $\boldsymbol{\cal P}_2=k^{-1/2}\boldsymbol{\cal P}_2^{(2)}$ and $\boldsymbol{\cal Q}_2=k^{-1/2}\boldsymbol{\cal Q}_2^{(2)}$ when $k \ge k_2$ and for $\boldsymbol{\cal P}_2=\boldsymbol{\cal P}_2^{(0)}$ and $\boldsymbol{\cal Q}_2=\boldsymbol{\cal Q}_2^{(0)}$ when $k < k_2$, in which the complex vectors $\boldsymbol{\cal P}_2^{(0)}$, $\boldsymbol{\cal Q}_2^{(0)}$, $\boldsymbol{\cal P}_2^{(2)}$ and $\boldsymbol{\cal Q}_2^{(2)}$ are defined by equations~(138)-(146) of~\cite{Ardavan2021} and $k_2$ designates a threshold frequency.  

The variable $\sigma_{21}$ determines the separation between two nearby stationary points of the phases of the received waves: the smaller the value of $\sigma_{21}$, the more focused is the observed radiation and the higher is its frequency content~\citep[][Section~4.5]{Ardavan2021}.  The radio component of the present radiation is mostly generated by values of $\sigma_{21}$ that range from $10^{-4}$ to $10^{-2}$.  In this paper, we replace $\boldsymbol{\cal P}_2^{(0)}$, $\boldsymbol{\cal Q}_2^{(0)}$, $\boldsymbol{\cal P}_2^{(2)}$ and $\boldsymbol{\cal Q}_2^{(2)}$, which only have weak dependences on $\sigma_{21}$, by their values for $\sigma_{21}=10^{-3}$ and treat them as constant parameters.

Evaluation of the right-hand side of equation~(\ref{E2}) results in
\begin{eqnarray}
S_\nu&=&\kappa_1\, k^{-2/3-j/2}\Big[{\rm Ai}^2(-k^{2/3}\sigma_{21}^2)+\zeta_1^2k^{-2/3}{{\rm Ai}^\prime}^2(-k^{2/3}\sigma_{21}^2)\nonumber\\*
&&+2\zeta_1\cos\beta\, k^{-1/3}{\rm Ai}(-k^{2/3}\sigma_{21}^2){\rm Ai}^\prime(-k^{2/3}\sigma_{21}^2)\Big],
\label{E2}
\end{eqnarray}
where $j=0$ when $k<k_2$ and $j=2$ when $k\ge k_2$,
\begin{equation}
\kappa_1=\kappa_0\left\vert\boldsymbol{\cal P}^{(j)}_2\right\vert^2,\,\,\, \zeta_1=\frac{\left\vert{\boldsymbol{\cal Q}^{(j)}_2}\right\vert}{\left\vert{\boldsymbol{\cal P}^{(j)}_2}\right\vert},\,\,\,\cos\beta=\frac{\Im\left(\boldsymbol{\cal Q}^{(j)}_2\cdot\boldsymbol{\cal P}^{(j)*}_2\right)}{\left\vert\boldsymbol{\cal Q}^{(j)}_2\right\vert \left\vert\boldsymbol{\cal P}^{(j)}_2\right\vert},
\label{E3}
\end{equation}
and $\Im{}$ and $*$ denote an imaginary part and the complex conjugate, respectively.  The above spectrum is emblematic of any radiation that entails caustics~\citep[see][]{Stamnes1986}.

To take account of the fact that the parameter $\sigma_{21}$ assumes a non-zero range of values across the (non-zero) latitudinal width of the detected radiation beam~\citep[][Section~4.5]{Ardavan2021}, we must integrate $S_\nu$ with respect to $\sigma_{21}$ over a finite interval $\rho\sigma_0\le\sigma_{21}\le\sigma_0$ with $\sigma_0\ll1$ and $0\le\rho<1$.  Performing the integration of the Airy functions in equation~(\ref{E2}) with respect to $\sigma_{21}$ by means of Mathematica, we thus obtain 
\begin{eqnarray}
{\cal F}_\nu&=&\int_{\rho\sigma_0}^{\sigma_0}S_\nu\,{\rm d}\sigma_{21}\nonumber\\*
&=& \kappa\,\chi^{-j/2}\left[f_1(\chi,\rho)+\frac{\zeta^2}{4\sqrt{3}}f_2(\chi,\rho)-\frac{\zeta\cos\beta}{2\sqrt{3}}f_3(\chi,\rho)\right],\nonumber\\*
\label{E4}
\end{eqnarray}
where
\begin{equation}
\kappa=\frac{2^{j/2}\sigma_0^{3(1+j/2)}}{3^{(j+3)/2}\pi^{3/2}}\kappa_1,\qquad \zeta=\sigma_0 \zeta_1,\qquad\chi=\frac{2}{3}{\sigma_0}^3k,
\label{E5}
\end{equation}
\begin{eqnarray}
f_1&=&\bigg[3\Gamma\left(\frac{7}{6}\right)\eta\chi^{-2/3}{}_2F_3
\left(\begin{matrix}
1/6&1/6&{}\\
1/3&2/3&7/6
\end{matrix}
;-\eta^6\chi^2
\right)\nonumber\\
&&+\pi^{1/2}\eta^3{}_2F_3
\left(\begin{matrix}
1/2&1/2&{}\\
2/3&4/3&3/2
\end{matrix}
\,;-\eta^6\chi^2
\right)\nonumber\\
&&+\frac{9}{20}\Gamma\left(\frac{5}{6}\right)\eta^5\chi^{2/3}{}_2F_3
\left(\begin{matrix}
5/6&5/6&{}\\
4/3&5/3&11/6
\end{matrix}
\,;-\eta^6\chi^2
\right)\bigg]_{\eta=\rho}^{\eta=1},\nonumber\\
\label{E6}
\end{eqnarray}
\begin{equation}
f_2=\eta\chi^{-4/3}\,{}_{24}G^{31}\left(-\eta^2\chi^{2/3},\frac{1}{3}\left\vert\,\begin{matrix}
5/6&7/6&{}&{}\\
0&2/3&4/3&-1/6
\end{matrix}\right)\right.\bigg\vert_{\eta=\rho}^{\eta=1},
\label{E7}
\end{equation}
\begin{equation}
f_3=\eta\chi^{-1}\,{}_{24}G^{31}\left(-\eta^2\chi^{2/3},\frac{1}{3}\left\vert\,\begin{matrix}
5/6&1/2&{}&{}\\
0&1/3&2/3&-1/6
\end{matrix}\right)\right.\bigg\vert_{\eta=\rho}^{\eta=1},
\label{E8}
\end{equation}
and ${}_2F_3$ and ${}_{24}G^{31}$ are respectively the generalised hypergeometric function~\citep[see][]{Olver} and the generalised Meijer G-Function \footnote{{\url{https://mathworld.wolfram.com/Meijer G-Function.html}}}.  The variable $\chi$ that appears in the above expressions is related to the frequency $\nu$ of the radiation via $\chi=4\pi{\sigma_0}^3\nu/(3\omega)$.

The scale and shape of the spectrum described by equation~(\ref{E4}) depend on whether $j$ equals $0$ or $2$ (i.e. on whether the dimensionless frequency $k$ lies below or above the threshold frequency $k_2$) and on the five parameters $\kappa$, $\zeta$, $\beta$, $\sigma_0$ and $\rho$: parameters whose values are dictated by the characteristics of the magnetospheric current sheet (see Section~\ref{sec:connection}).  The parameters $\zeta$, $\beta$  and $\rho$ determine the shape of the spectral distribution while the parameters $\kappa$ and $\sigma_0$ respectively determine the position of this distribution along the flux-density (${\cal F}_\nu$) and the frequency ($\nu$) axes.

\section{Fits to the data on examples of various forms of radio spectra}
\label{sec:fits}

In this section, we choose from each of the five galleries of pulsar spectral shapes catalogued by~\citet{Swainston} two examples with the largest number of known data points and fit them with the spectral distribution function described by equation~(\ref{E4}).  In the case of each example, we use Mathematica's {\tt `NonlinearModelFit'} procedure\footnote{ {\url{https://reference.wolfram.com/language/ref/NonlinearModelFit.html}}} and the statistical information that it provides to determine the values of the fit parameters in equation~(\ref{E4}) and their standard errors.  Where, owing to the complexity of the expression in equation~(\ref{E4}), this procedure fails to work and so the fits to the data are obtained by elementary iteration, only the values of these parameters are specified.  

\begin{figure}
\centerline{\includegraphics[width=9.75cm]{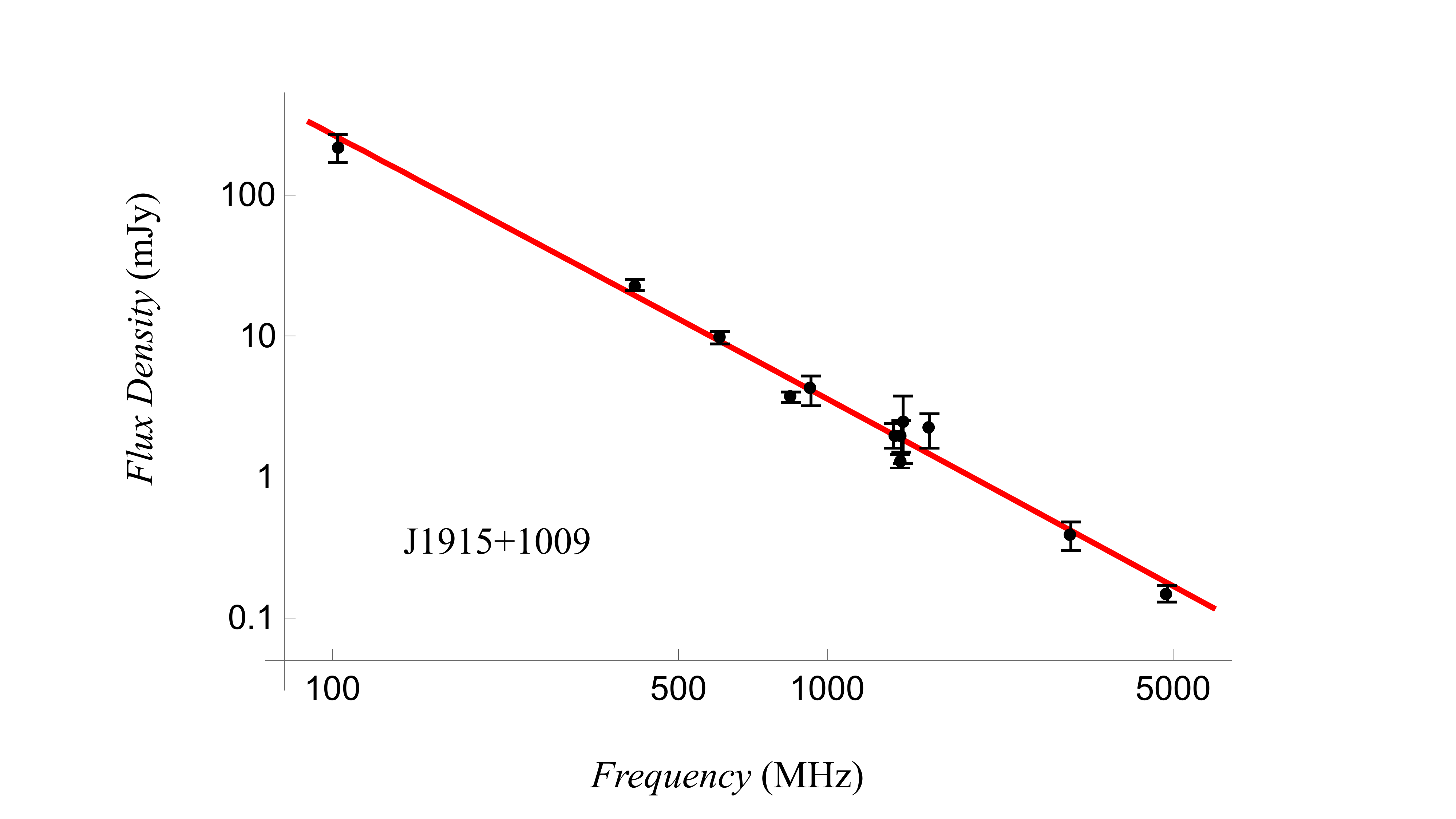}}
\caption{Spectrum of J1915+1009: an example from the simple-power-law gallery of~\citet{Swainston}.  The curve is a plot of the flux density ${\cal F}_\nu$ given by equation~(\ref{E4}) for $j=2$, $\kappa=8.39\times10^3$ mJy,  $\sigma_0=8.22\times10^{-3}$, $\zeta=\beta=\rho=0$ and $\chi=0.150\,\nu_{\rm MHz}$.}
\label{RadioF1}
\end{figure}

\begin{figure}
\centerline{\includegraphics[width=9.75cm]{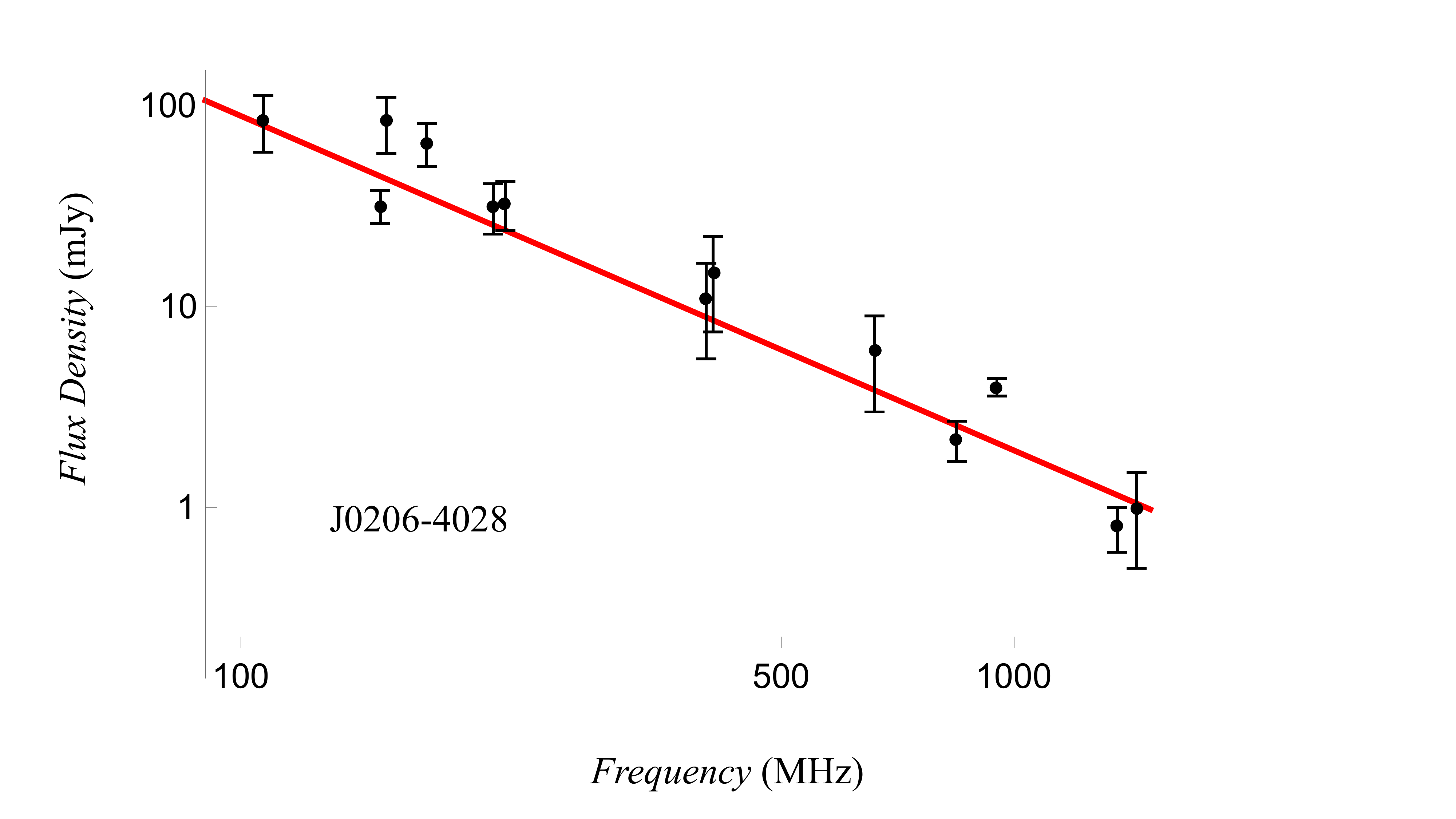}}
\caption{Spectrum of J0206-4028: a second example from the simple-power-law gallery of~\citet{Swainston}.  The curve is a plot of the flux density ${\cal F}_\nu$ given by equation~(\ref{E4}) for $j=2$, $\kappa=2.58\times10^3$ mJy,  $\sigma_0=7.75\times10^{-3}$, $\zeta=0.668$, $\beta=\rho=0$ and $\chi=0.112\,\nu_{\rm MHz}$.}
\label{RadioF2}
\end{figure}

\begin{figure}
\centerline{\includegraphics[width=10cm]{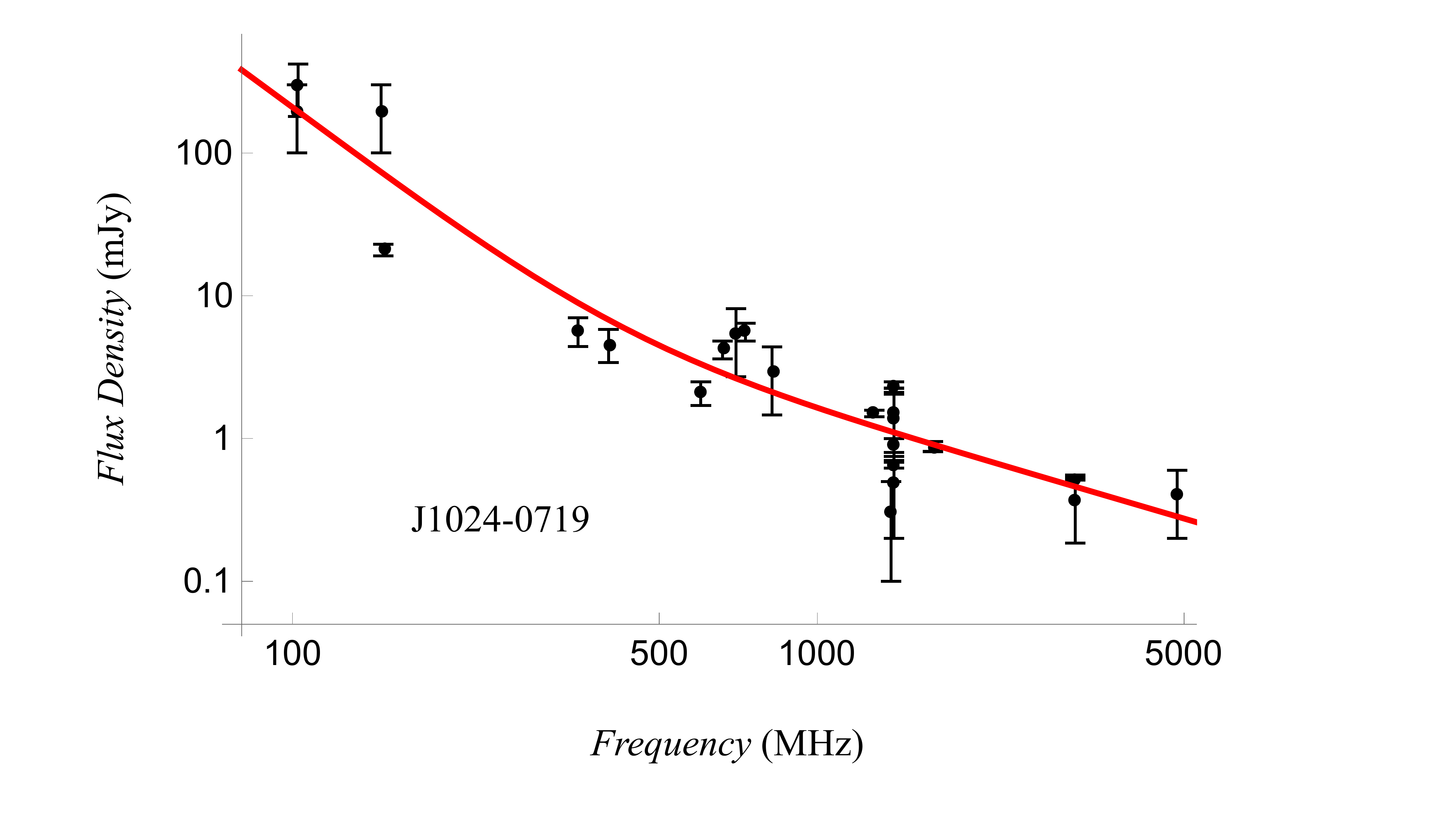}}
\caption{Spectrum of J1024-0719: an example from the broken-power-law gallery of~\citet{Swainston}.  The curve is a plot of the flux density ${\cal F}_\nu$ given by equation~(\ref{E4}) for $j=2$, $\kappa=2.81\times10^{-2}$ mJy,  $\sigma_0=2.81\times10^{-3}$, $\zeta=~0.507$, $\beta=0.5$, $\rho=0$ and $\chi=7.67\times10^{-5}\,\nu_{\rm MHz}$.}
\label{RadioF3}
\end{figure}

\begin{figure}
\centerline{\includegraphics[width=9.75cm]{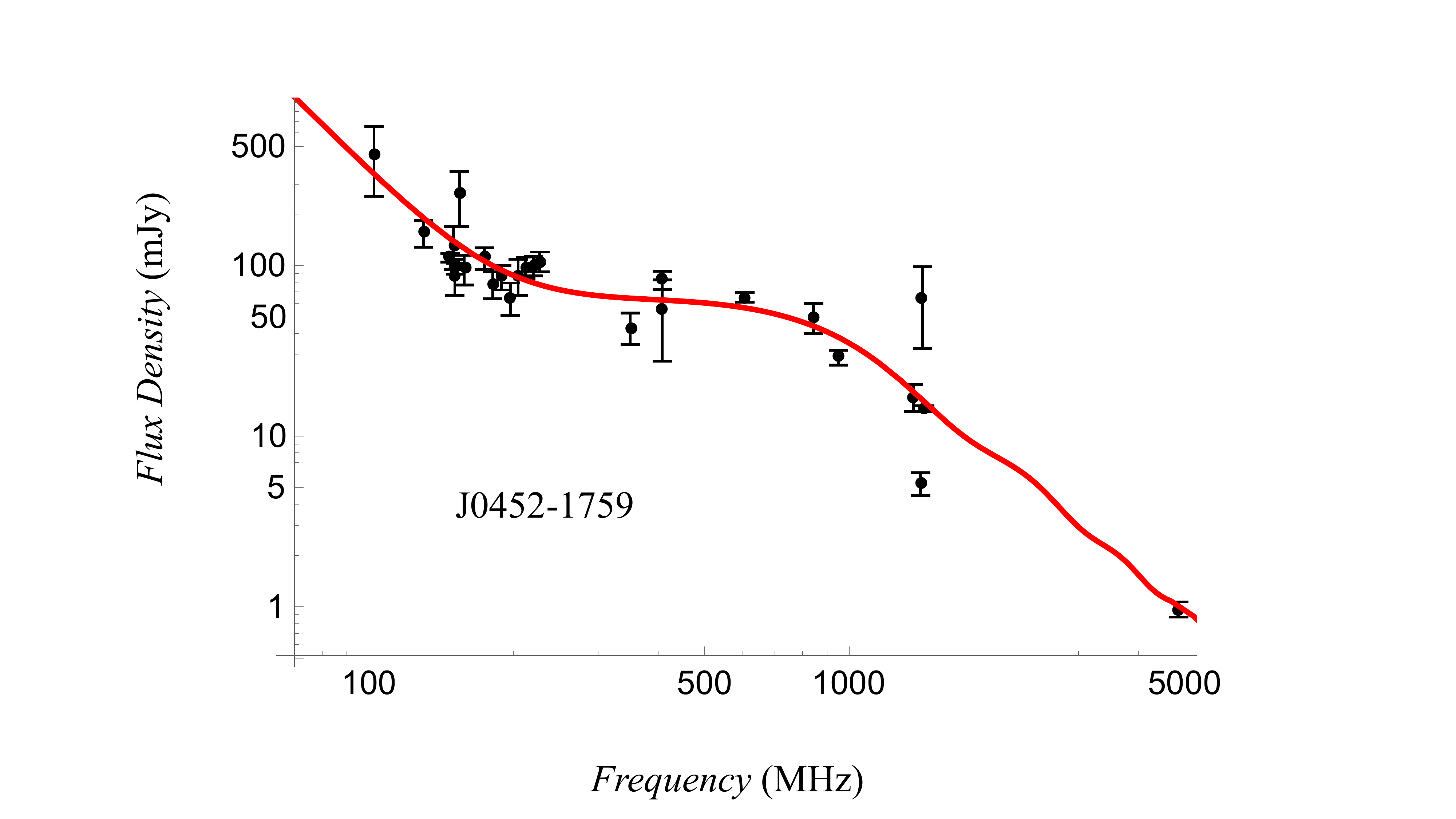}}
\caption{Spectrum of J0452-1759: a second example from the broken-power-law gallery of~\citet{Swainston}.  The curve is a plot of the flux density ${\cal F}_\nu$ given by equation~(\ref{E4}) for $j=2$, $\kappa=30.6\pm10.6$ mJy,  $\sigma_0=(1.88\pm0.18)\times10^{-3}$, $\zeta=1.75\pm0.59$, $\beta=0.64\pm0.18$, $\rho=0.58\pm0.34$ and $\chi=(2.42\pm0.62)\times10^{-3}\nu_{\rm MHz}$.}
\label{RadioF4}
\end{figure}

\begin{figure}
\centerline{\includegraphics[width=9.75cm]{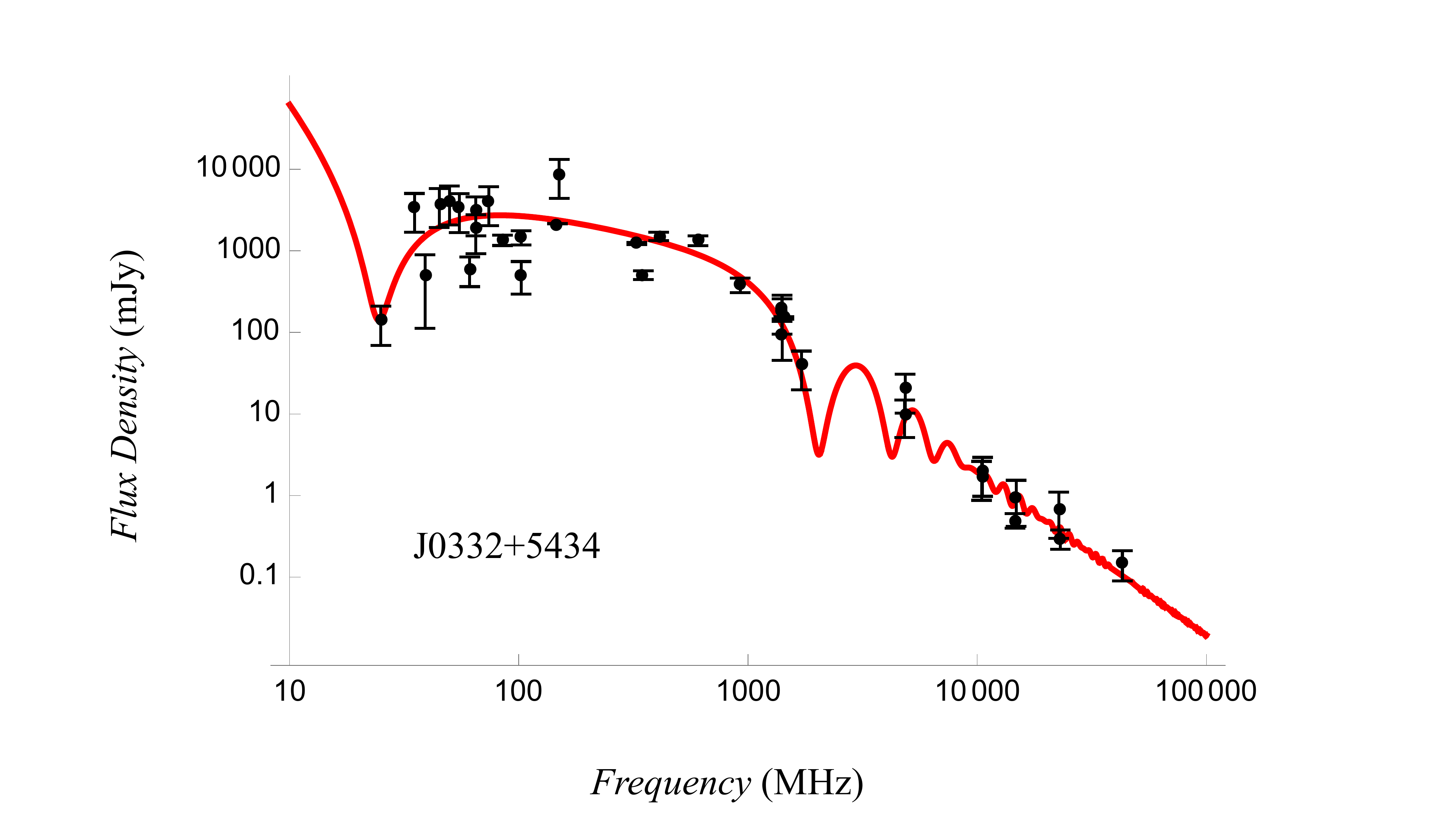}}
\caption{Spectrum of J0332+5434: an example from the low-frequency-turn-over gallery of~\citet{Swainston}.  The curve is a plot of the flux density ${\cal F}_\nu$ given by equation~(\ref{E4}) for $j=2$, $\kappa=(1.55\pm1.42)\times10^3$ mJy,  $\sigma_0=(1.49\pm0.06)\times10^{-3}$, $\zeta=0.592\pm0.033$, $\beta=(3.99\pm2.00)\times10^{-2}$, $\rho=0.927\pm0.077$ and $\chi=(1.58\pm0.20)\times10^{-3}\,\nu_{\rm MHz}$.}
\label{RadioF5}
\end{figure}

\begin{figure}
\centerline{\includegraphics[width=9.75cm]{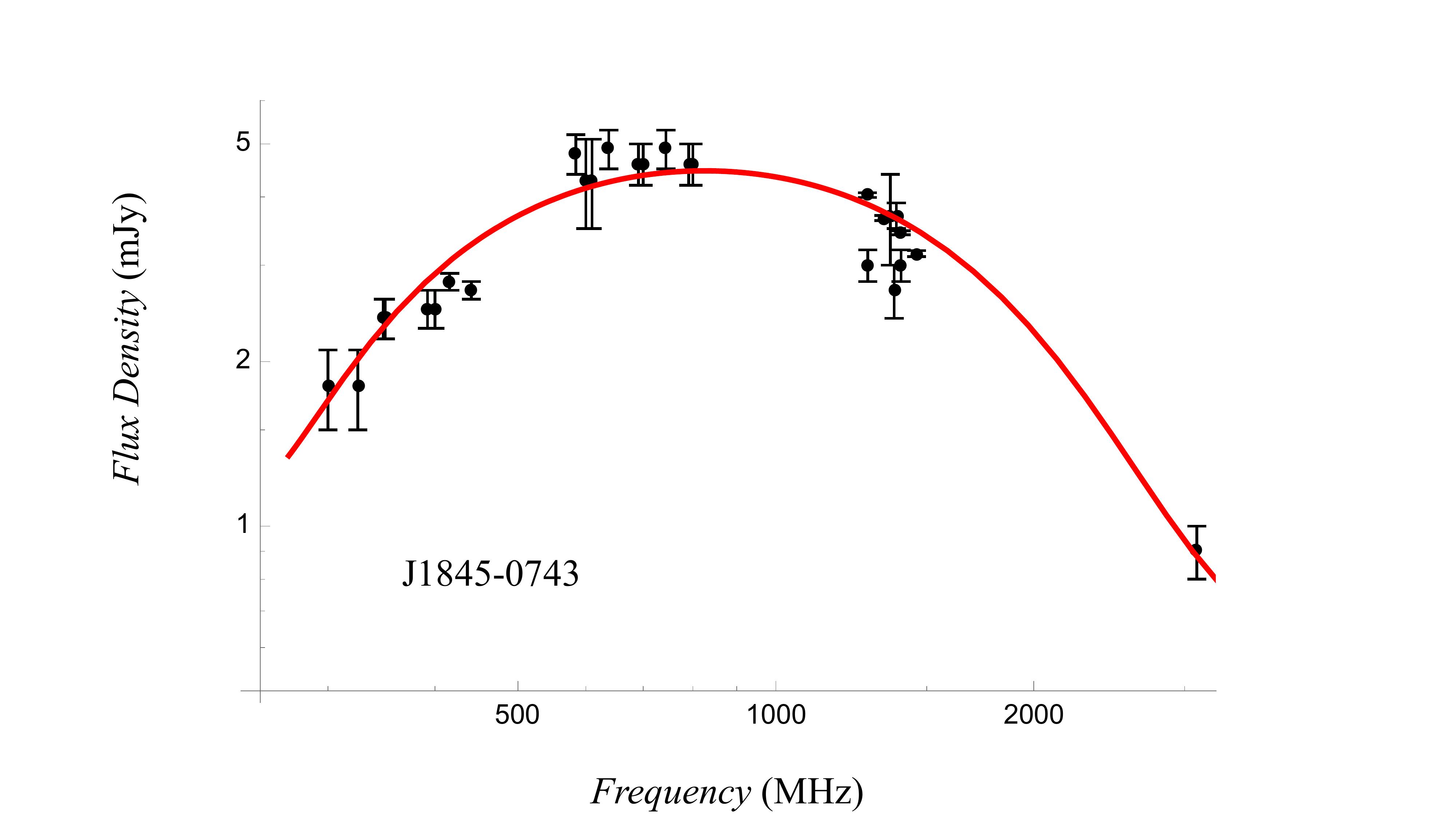}}
\caption{Spectrum of J1845-0743: a second example from the low-frequency-turn-over gallery of~\citet{Swainston}.  The curve is a plot of the flux density ${\cal F}_\nu$ given by equation~(\ref{E4}) for $\kappa=2.22\pm0.12$ mJy,  $\sigma_0=(2.94\pm0.08)\times10^{-3}$, $\zeta=2.08\pm0.29$, $\beta=0$, $\rho=0.418\pm0.056$ and $\chi=(1.77\pm0.14)\times10^{-3}\,\nu_{\rm MHz}$.}
\label{RadioF6}
\end{figure}

\begin{figure}
\centerline{\includegraphics[width=9.75cm]{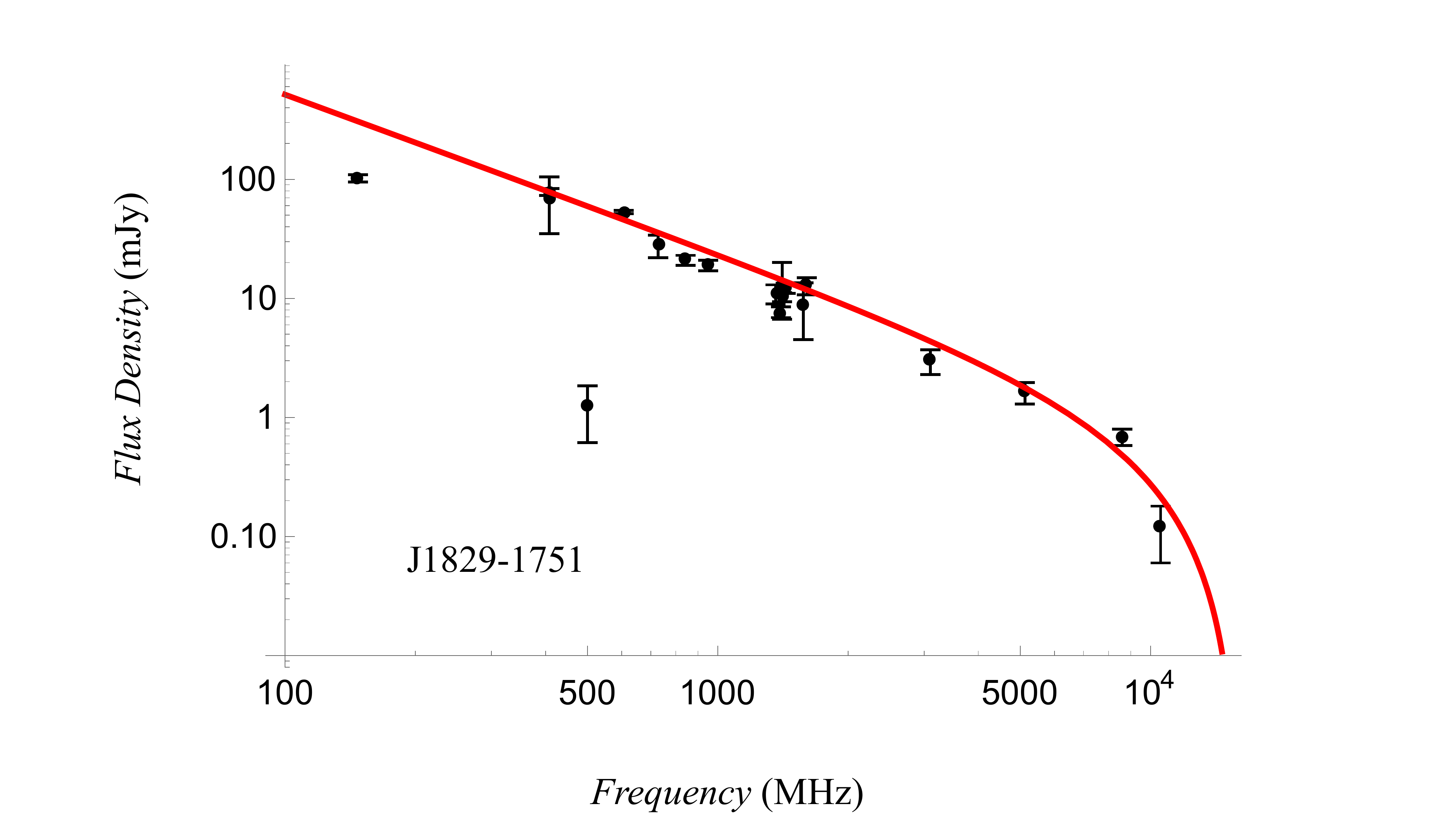}}
\caption{Spectrum of J1829-1751; an example from the high-frequency-cut-off gallery of~\citet{Swainston}.  The curve is a plot of the flux density ${\cal F}_\nu$ given by equation~(\ref{E4}) for $j=0$, $\kappa=3.16\times10^{-4}$ mJy,  $\sigma_0=5.93\times10^{-4}$, $\zeta=10^3$, $\beta=0$, $\rho=0.999$ and $\chi=4.27\times10^{-5}\,\nu_{\rm MHz}$.}
\label{RadioF7}
\end{figure}

\begin{figure}
\centerline{\includegraphics[width=9.75cm]{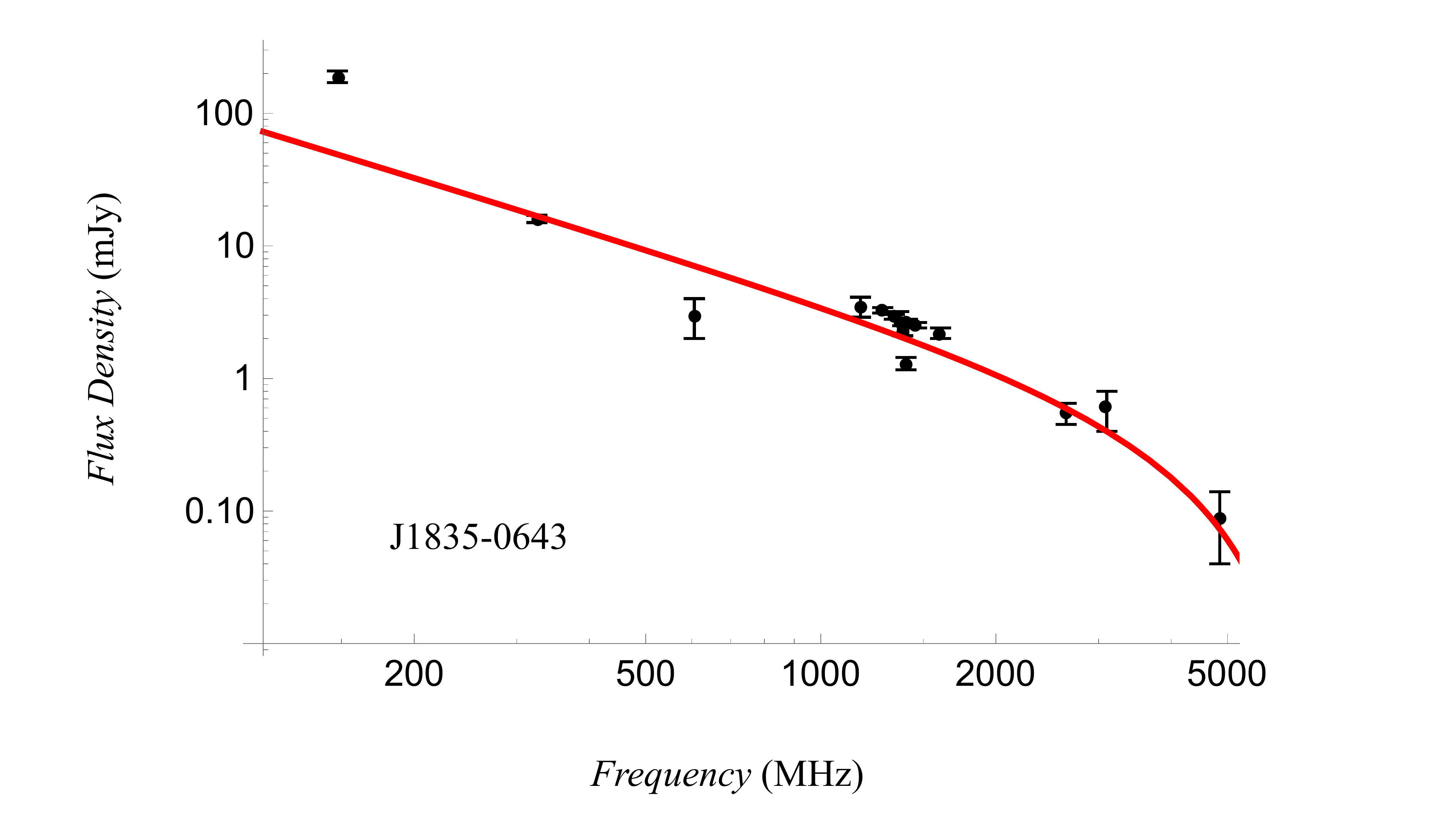}}
\caption{Spectrum of J1835-0643: a second example from the high-frequency-cut-off gallery of~\citet{Swainston}.  The curve is a plot of the flux density ${\cal F}_\nu$ given by equation~(\ref{E4}) for $j=0$, $\kappa=1.58\times10^{-4}$ mJy,  $\sigma_0=7.88\times10^{-4}$, $\zeta=10^3$, $\beta=0$, $\rho=0.999$ and $\chi=10^{-4}\,\nu_{\rm MHz}$.}
\label{RadioF8}
\end{figure}

\begin{figure}
\centerline{\includegraphics[width=9.75cm]{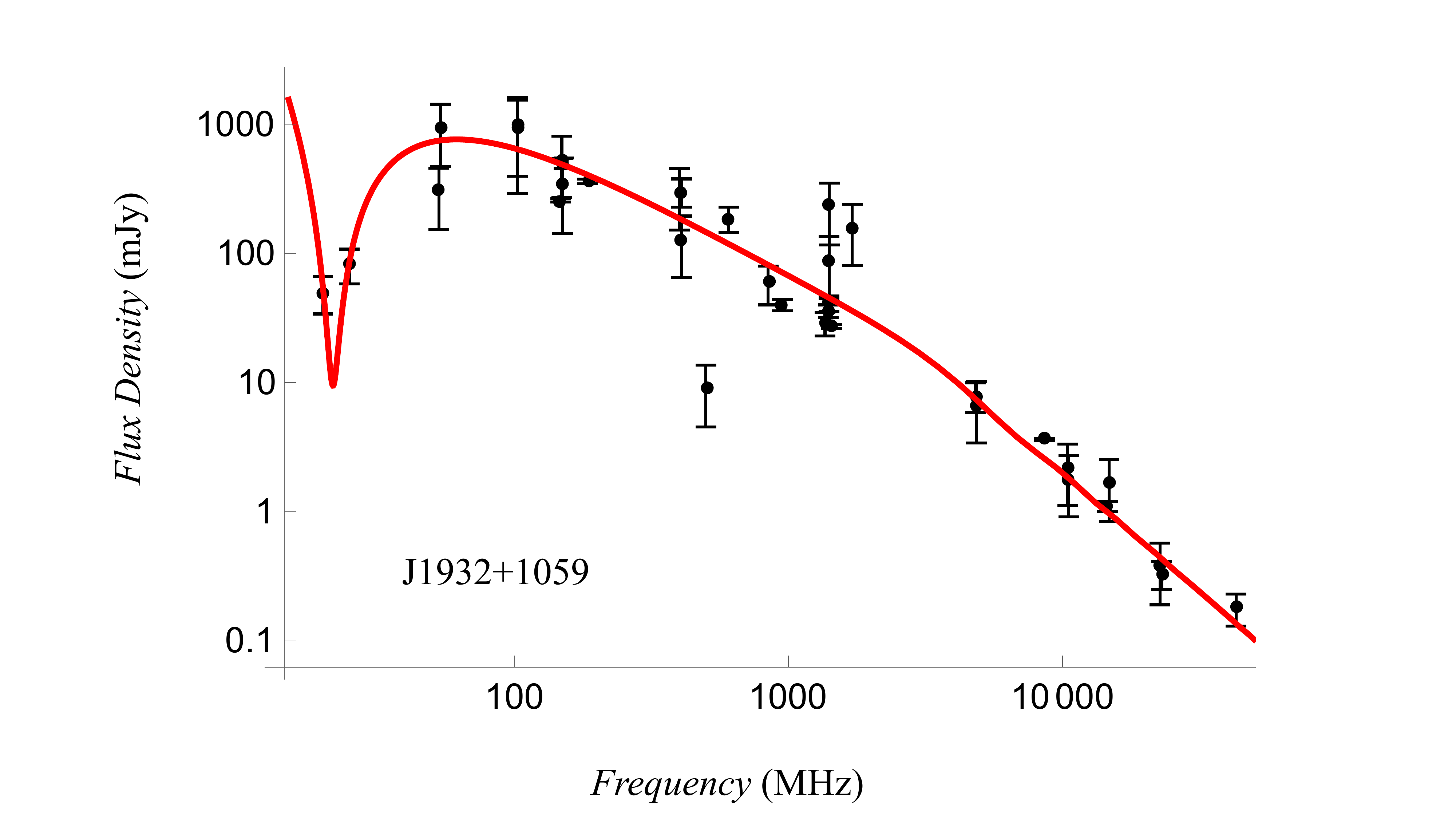}}
\caption{Spectrum of J1932+1059: an example from the double-turn-over-spectrum gallery of~\citet{Swainston}.  The curve is a plot of the flux density ${\cal F}_\nu$ given by equation~(\ref{E4}) for $j=2$, $\kappa=8.38\pm7.35$ mJy,  $\sigma_0=(1.49\pm0.36)\times10^{-3}$, $\zeta=0.354\pm0.069$, $\beta=\rho=0$ and $\chi=(5.02\pm2.84)\times10^{-4}\,\nu_{\rm MHz}$.}
\label{RadioF9}
\end{figure}

\begin{figure}
\centerline{\includegraphics[width=9.75cm]{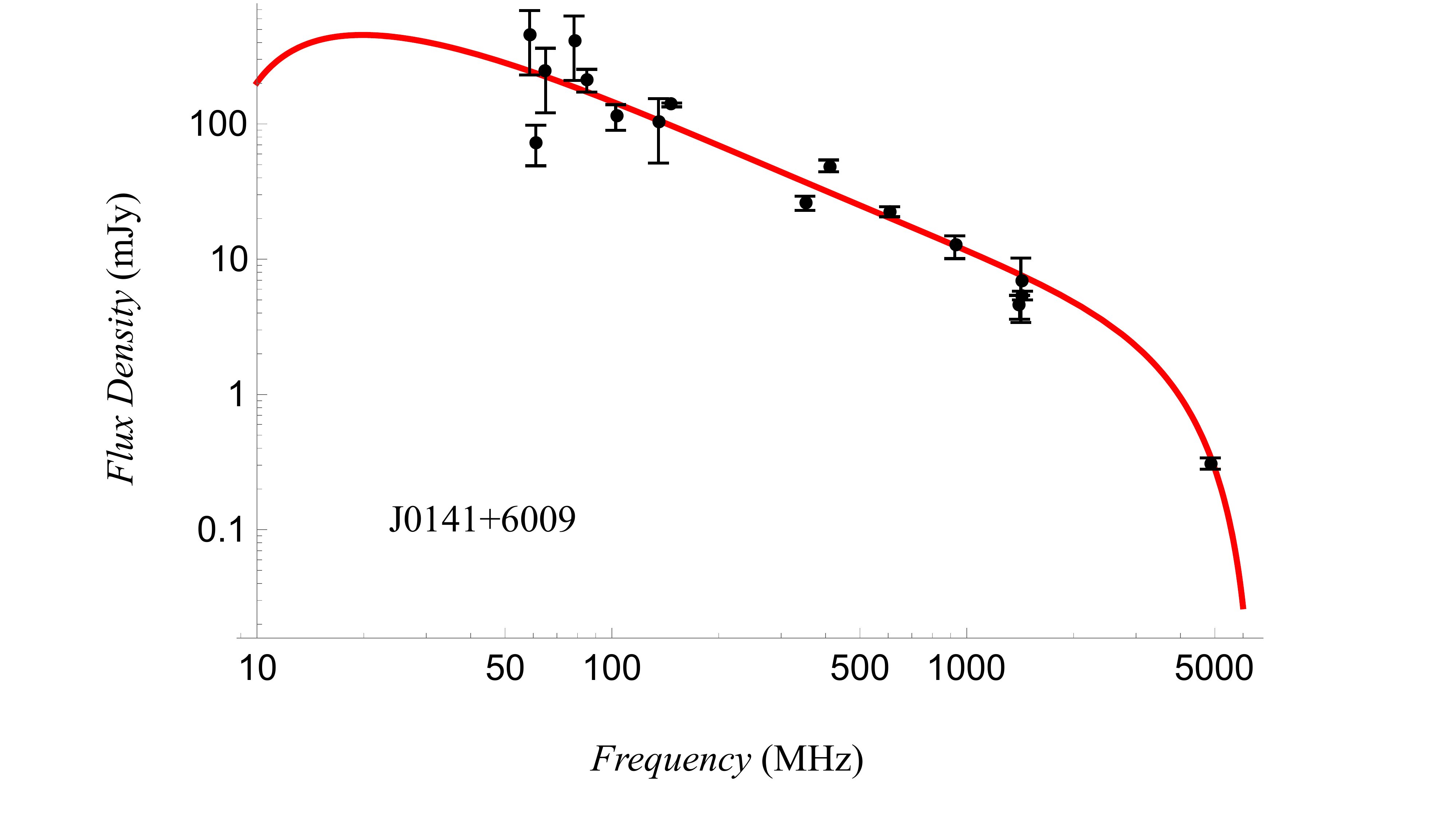}}
\caption{Spectrum of J0141+6009: a second example from the double-turn-over-spectrum gallery of~\citet{Swainston}.  The curve is a plot of the flux density ${\cal F}_\nu$ given by equation~(\ref{E4}) for $j=2$, $\kappa=5.01\times~10^2$ mJy,  $\sigma_0=7.88\times10^{-4}$, $\zeta=0.225$, $\beta=0$, $\rho=0.999$ and $\chi=3.98\times~10^{-4}\,\nu_{\rm MHz}$.}
\label{RadioF10}
\end{figure}

The results are presented in Figs.~(\ref{RadioF1})--(\ref{RadioF10}) and their captions.  The horizontal and vertical axes in these logarithmic plots are marked with the values of ${\cal F}_\nu$ and $\nu_{\rm MHz}$, respectively, where $\nu_{\rm MHz}$ stands for the frequency of the radiation in units of MHz.  

It can be seen from Figs.~(\ref{RadioF1})--(\ref{RadioF10}) that the fit residuals in the case of each pulsar are smaller than the corresponding observational errors for the majority of the data points.  Since there are no two values of any of the fit parameters for which the spectrum described by equation~(\ref{E4}) has the same shape and position, the specified values of the fit parameters are moreover unique.

\section{The connection between the parameters of a fitted spectrum and the physical characteristics of the source of the observed radiation}
\label{sec:connection}

\subsection{Derivation of the connecting relations}
\label{subsec:connecting}

From equations~(\ref{E5}), (\ref{E3}) and (\ref{E1}), it follows that the parameters $\kappa$, $\zeta$, $\rho$ and $\sigma_0$ in equation~(\ref{E4}) are related to the characteristics of the source of the observed radiation via the quantities $\omega$, $\kappa_0$, $\boldsymbol{\cal P}_2^{(j)}$ and $\boldsymbol{\cal Q}_2^{(j)}$ that appear in the expression for the flux density ${\cal F}_\nu$.  In this section we use the results of the analysis presented in~\citet{Ardavan2021} to express these quantities in terms of the inclination angle of the central neutron star, $\alpha$, the magnitude of the star's magnetic field at its magnetic pole, $B_0=10^{12}{\hat B}_0$ Gauss, the radius of the star, $r_{s0}=10^6 d$ cm, the rotation frequency of the star, $\omega=10^2{\hat P}^{-1}$ rad/s, and the spherical polar coordinates, $R_P=D\,\,{\rm kpc}=3.085\times10^{21}D$~cm, $\varphi_P$ and $\theta_P$, of the observation point $P$ in a frame whose centre and $z$-axis coincide with the centre and spin axis of the star.

From equations~(177), (138)--(146) and (136) of~\citet{Ardavan2021} it follows that
\begin{equation}
\kappa_0=4.15\times10^{18} w_1^2 w_3^2 {\hat B}_0^2 d^4 D^{-2} {\hat P}^{-1}\quad {\rm Jy},
\label{E9}
\end{equation}

\begin{equation}
\left\vert\boldsymbol{\cal P}_2^{(0)}\right\vert=\frac{b}{6}\left\vert{\tilde{\bf P}}_2\right\vert\left(\sqrt{\frac{2\sigma_{21}}{\partial^2 f_{2C}/\partial\tau^2|_{\tau=\tau_{0{\rm min}}}}}+\sqrt{\frac{-2\sigma_{21}}{\partial^2 f_{2C}/\partial\tau^2|_{\tau=\tau_{0{\rm max}}}}}\right),
\label{E10}
\end{equation}
and
\begin{equation}
\left\vert\boldsymbol{\cal P}_2^{(2)}\right\vert=\left(\frac{2}{\pi a^3}\right)^{1/2}\left\vert\boldsymbol{\cal P}_2^{(0)}\right\vert,
\label{E11}
\end{equation}
where
\begin{equation}
w_1=\left\vert1-2\alpha/\pi\right\vert,\qquad w_3=1+0.2\sin^2\alpha,
\label{E12}
\end{equation}
\begin{eqnarray}
f_{2C}&=&({\hat r}_P^2{\hat r}_{sC}^2\sin^2\theta-1)^{1/2}-{\hat R}_P-\arccos\left({\hat r}_P^{-1}{\hat r}_{sC}^{-1}\csc\theta\right)\nonumber\\*
&&-\arccos(\cot\alpha\cot\theta)+{\hat r}_{sC}+\varphi_P-{\hat r}_{s0}-2\pi,
\label{E13}
\end{eqnarray}
\begin{eqnarray}
a&=&({\hat r}_P^2{\hat r}_{sC}^2\sin^2\theta-1)\left[\frac{({\hat r}_P^2{\hat r}_{sC}^2\sin^2\theta-1)^{1/2}+{\hat r}_{sC}}{{\hat r}_{sC}({\hat r}_P^2-1)^{1/2}({\hat R}_P^2\sin^2\theta-1)^{1/2}}\right.\nonumber\\*
&&\left.-\frac{1}{({\hat r}_P^2{\hat r}_{sC}^2\sin^2\theta-1)^{1/2}}\right],
\label{E14}
\end{eqnarray}
\begin{equation}
b=\frac{{\hat r}_P^2{\hat r}_{sC}^2\sin^2\theta-1}{{\hat r}_{sC}({\hat r}_P^2-1)^{1/2}({\hat R}_P^2\sin^2\theta-1)^{1/2}},
\label{E15}
\end{equation}
\begin{equation}
{\hat r}_{sC}=\frac{({\hat r}_P^2-1)^{1/2}({\hat R}_P^2\sin^2\theta-1)^{1/2}-{\hat z}_P\cos\theta}{{\hat r}_P^2\sin^2\theta-1},
\label{E16}
\end{equation}
\begin{equation}
\sigma_{21}=[{\textstyle\frac{3}{4}}(f_{2C}\vert_{\tau=\tau_{2{\rm max}}}-f_{2C}\vert_{\tau=\tau_{2{\rm min}}})]^{1/3},
\label{E17}
\end{equation}
\begin{equation}
\theta=\arccos(\sin\alpha\cos\tau),
\label{E18}
\end{equation}
\begin{eqnarray}
\left\vert{\tilde{\bf P}}_2\right\vert &=&\cos\alpha[{\hat r}_{sC}^2(1+\cos^2\theta_P-2\cos\theta\cos\theta_P)+\tan^2\alpha\nonumber\\*
&&-\cot^2\theta+2{\hat r}_{sC}\sin\theta\cos\theta_P(\tan^2\alpha-\cot^2\theta)^{1/2}]^{1/2},\quad\,\,\nonumber\\*
\label{E19}
\end{eqnarray}
and the variables $\tau_{2{\rm min}}$ and $\tau_{2{\rm max}}$ stand for the minimum and maximum of the function $f_{2C}$ (see also equations~7, 9, 174, 175, 93--95, 97 and 88 of~\citealt{Ardavan2021}).  The caret on $R_P$ and $r_{s0}$ (and $r_P=R_P\sin\theta_P$, $z_P=R_P\cos\theta_P$) is used here to designate a variable that is rendered dimensionless by being measured in units of the light-cylinder radius $c/\omega$.  (Note the following two corrections: the vector ${\bf P}_2$ in equation~145 and the numerical coefficient $1.54\times10^{19}$ in equation~177 of~\citealt{Ardavan2021} have been corrected to read ${\tilde{\bf P}}_2$ and $4.15\times10^{18}$, respectively.)

The expression for $\vert{\tilde{\bf P}}_2\vert$ in equation~(\ref{E19}) is derived from that for ${\bf P}_2$ given by equations~(98), (80), (78) and (62) of~\citet{Ardavan2021}.  In this derivation, we have set the observation point on the cusp locus of the bifurcation surface where ${\hat r}_s={\hat r}_{sC}$, have approximated $(p_1,p_2,p_3)$ by its far-field value $2^{1/3}{\hat R}_P^{-1}({\hat R}_P^{-1}, -1, 1)$ and have let ${\bf P}_2={\hat R}_P^{-1}{\tilde{\bf P}}_2$.  The factor ${\hat R}_P^{-2}$ that would have otherwise appeared in the resulting expression for $\vert\boldsymbol{\cal P}^{(2)}_2\vert^2$ is thus incorporated in the coefficient $\kappa_0$ in equation~(\ref{E9}).

\begin{figure}
\centerline{\includegraphics[width=10cm]{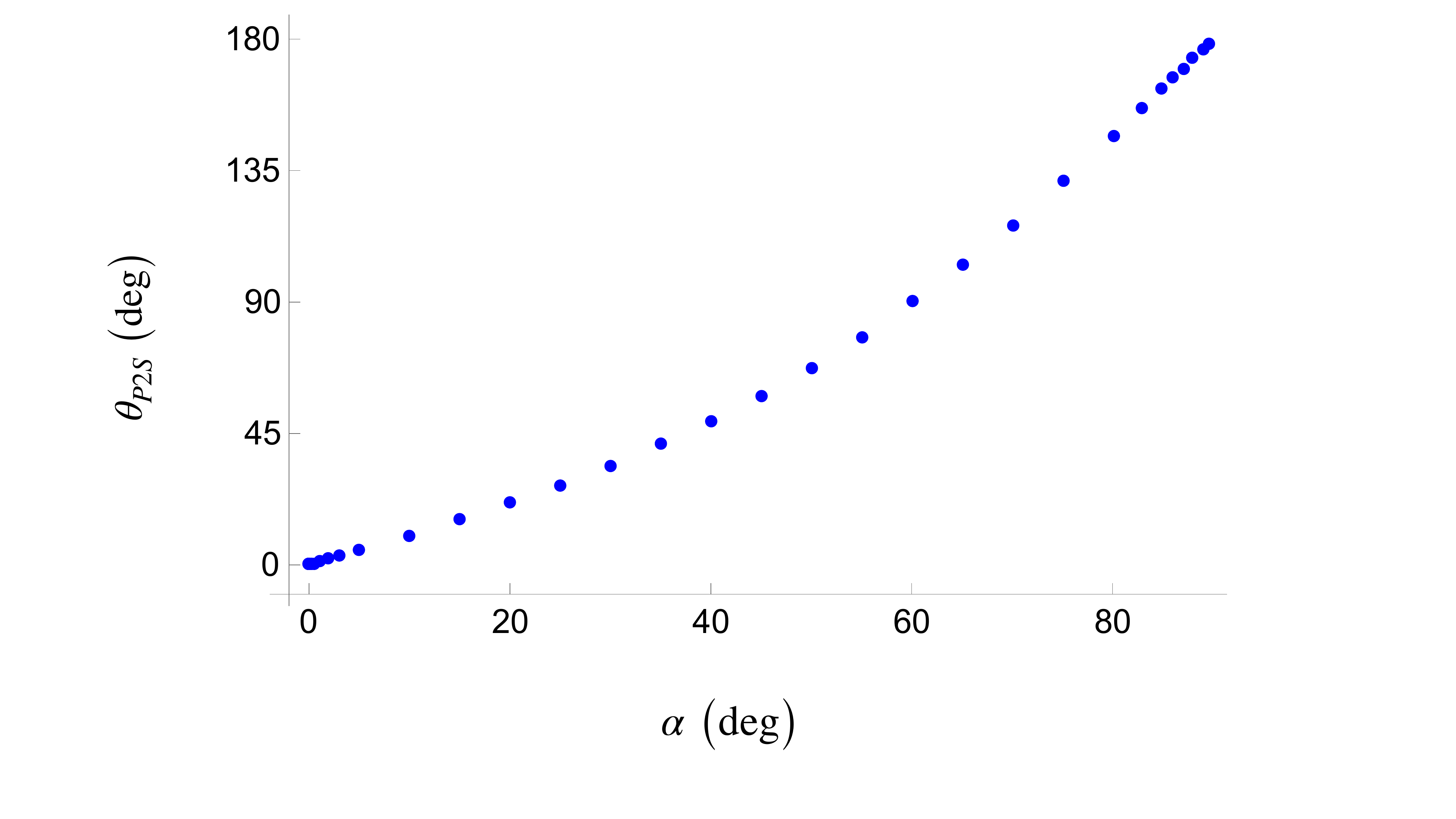}}
\caption{The relationship between the colatitude $\theta_{P2S}$ on which the focused radiation beam is centred and the star's inclination angle $\alpha$.}
\label{RadioF11}
\end{figure}

For certain values of $\theta_P$, denoted by $\theta_{P2S}$, the function $f_{2C}(\tau)$ has an inflection point~(see~\citealt[][Section~4.4]{Ardavan2021}, and~\citealt[][Section~2]{Ardavan2023Crab}).  For any given inclination angle $\alpha$, the position $\tau_{2S}$ of this inflection point and the colatitude $\theta_{P2S}$ of the observation points for which $f_{2C}(\tau)$ has an inflection point follow from the solutions to the simultaneous equations $\partial f_{2C}/\partial\tau=0$ and $\partial^2 f_{2C}/\partial\tau^2=0$.  (Explicit expressions for the derivatives that appear in these equations can be found in Appendix A of~\citealt{Ardavan2021}.)   For values of $\theta_P$ sufficiently close to $\theta_{P2S}$, the separation between the maximum $\tau=\tau_{2{\rm max}}$ and minimum $\tau=\tau_{2{\rm min}}$ of $f_{2C}$ and hence the value of $\sigma_{21}$ are small.  In other words, the focused radiation beam is centred on the colatitude $\theta_{P2S}$ plotted in Fig.~\ref{RadioF11}~\citep[][Section~5.5]{Ardavan2021}.  Since the fit parameter $\sigma_0$, which denotes the maximum value assumed by $\sigma_{21}$, lies between $10^{-2}$ and $10^{-4}$ for most of the examples considered in Section~\ref{sec:fits}, we have chosen the separation between $\tau=\tau_{2{\rm max}}$ and $\tau=\tau_{2{\rm min}}$ in each example such that the value of $\sigma_{21}$ in equation~(\ref{E17}) is of the order of $10^{-3}$.  

Equations~(\ref{E5}), (\ref{E3}), and (\ref{E9})--(\ref{E11}) jointly yield
\begin{equation}
\kappa=1.59\times10^{16} \sigma_0^3{\hat B}_0^2d^4D^{-2}{\hat P}^{-1}{\tilde\kappa}_{\rm th}\qquad{\rm Jy},
\label{E20}
\end{equation}
with
\begin{eqnarray}
{\tilde\kappa}_{\rm th}&=&w_1^2 w_3^2 b^2\left\vert{\tilde{\bf P}}_2\right\vert^2\sigma_{21}\left(\frac{1} {\partial^2 f_{2C}/\partial\tau^2|_{\tau=\tau_{2{\rm min}}}}\right.\nonumber\\*
&&\left.-\frac{1} {\partial^2 f_{2C}/\partial\tau^2|_{\tau=\tau_{2{\rm max}}}}\right),
\label{E21}
\end{eqnarray}
when $j=0$, and
\begin{equation}
\kappa=6.76\times10^{15} \sigma_0^6{\hat B}_0^2d^4D^{-2}{\hat P}^{-1}{\hat\kappa}_{\rm th}\qquad{\rm Jy},
\label{E22}
\end{equation}
with
\begin{equation}
{\hat\kappa}_{\rm th}=a^{-3}{\tilde\kappa}_{\rm th},
\label{E23}
\end{equation}
when $j=2$.  Equations~(\ref{E20}) and (\ref{E22}) can be written as
\begin{equation}
 {\tilde\kappa}_{\rm th}={\tilde\kappa}_{\rm obs}\quad {\rm and}\quad{\hat\kappa}_{\rm th}={\hat\kappa}_{\rm obs},
 \label{E24}
 \end{equation}
respectively, where
\begin{equation}
{\tilde\kappa}_{\rm obs}=6.27\times10^{-17}\sigma_0^{-3}({\hat B}_0d^2)^{-2}D^2{\hat P}\kappa
\label{E25}
\end{equation}
and
\begin{equation}
{\hat\kappa}_{\rm obs}=1.48\times10^{-16}\sigma_0^{-6}({\hat B}_0d^2)^{-2}D^2{\hat P}\kappa.
\label{E26}
\end{equation}
While ${\tilde\kappa}_{\rm obs}$ and ${\hat\kappa}_{\rm obs}$ only contain the observed parameters of the pulsar and its emission, the values of ${\tilde\kappa}_{\rm th}$ and ${\hat\kappa}_{\rm th}$ are determined by the physical characteristics of the magnetospheric current sheet that acts as the source of the observed emission.  

\begin{figure}
\centerline{\includegraphics[width=9.75cm]{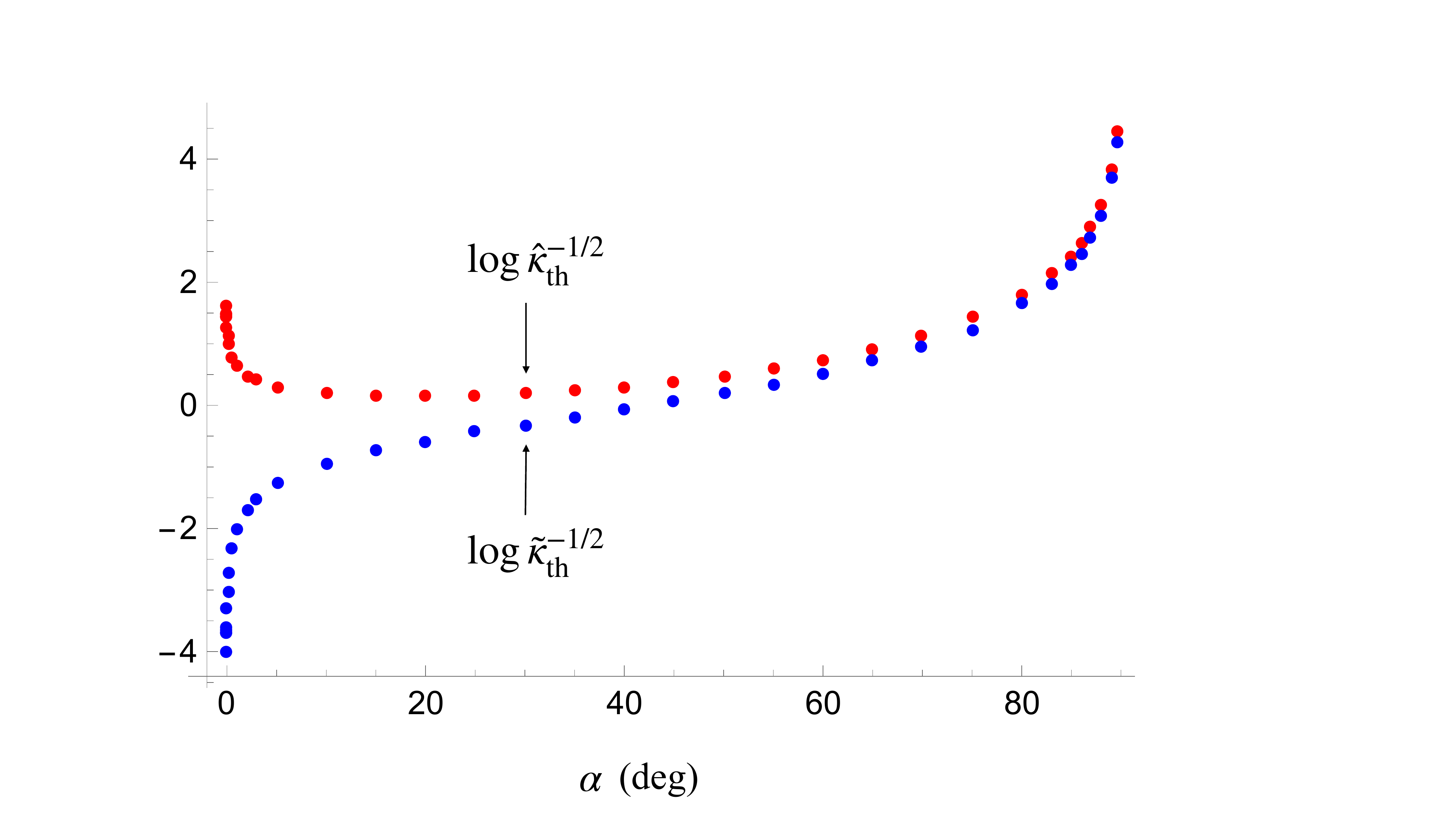}}
\caption{Plots of the dependences of the functions ${\hat\kappa}_{\rm th}$ and ${\tilde\kappa}_{\rm th}$, defined in equations~(\ref{E21}) and (\ref{E23}), on the inclination angle $\alpha$ for $\sigma_{21}\simeq10^{-3}$ and an observation point in the far zone.  The red dots and the blue dots depict the values assumed by ${\hat\kappa}_{\rm th}$ and ${\tilde\kappa}_{\rm th}$, respectively.}
\label{RadioF12}
\end{figure}

For a given value of $\sigma_{21}$, the right-hand sides of equations~(\ref{E21}) and (\ref{E23}) are functions of the inclination angle $\alpha$ and the observer's distance ${\hat R}_P$ only (see Fig.~\ref{RadioF12}).  The values of the fit parameters together with the relations in equation~(\ref{E24}) and the plots of ${\tilde\kappa}_{\rm th}$ and ${\hat\kappa}_{\rm th}$ in Fig.~\ref{RadioF12} thus enable us to connect the parameters of the fitted spectra to the physical characteristics of their sources.

\subsection{Application of the connecting relations to the fitted spectra}
\label{subsec:application}

In this section, we use the relations derived in Section~\ref{subsec:connecting}, the values of the fit parameters given in the captions to Figs.~\ref{RadioF1}--\ref{RadioF10} and the data listed in the ATNF Pulsar Catalogue~\citep{Manchester2005} to determine (or set limits on) certain attributes of the central neutron stars of the pulsars considered in Section~\ref{sec:fits} and their magnetospheres.

Once the values of the fit parameters $\sigma_0$ and $\kappa$ given in the caption to Fig.~\ref{RadioF1}, the period ($0.405$ s) and the distance ($7$ kpc) of the pulsar J1915+1009 are inserted in equation~(\ref{E26}), the resulting value of ${\hat\kappa}_{\rm obs}$ and the second member of equation~(\ref{E24}) yield
\begin{equation}
{\hat B}_0d^2=1.13\,{\hat\kappa}_{\rm th}^{-1/2}\quad{\rm for}\quad J1915+1009.
\label{E27}
\end{equation}
If the central neutron star of this pulsar has the radius $10^6$ cm (i.e. $d=1$) and the magnetic field $2.51\times10^{12}$ Gauss at its magnetic pole (i.e. ${\hat B}_0=2.51$) as predicted by the formula for magnetic dipole radiation, then equation~(\ref{E27}) and the curve delineated by the red dots in Fig.~\ref{RadioF12} imply that the value of the angle $\alpha$ between the rotation and magnetic axes of J1915+1009 is either $3.8^\circ$ or $42.3^\circ$.  (Here, and in other similar cases, a choice between the two possible values of $\alpha$ can be made by comparing the pulse profile of the pulsar in question with the theoretically predicted ones presented in Section~5.1 of~\citealt{Ardavan2021}.)  Depending on whether $\alpha=3.8^\circ$ or $42.3^\circ$, the direction along which the radiation is observed forms the angles $\theta_{P2S}=3.8^\circ$ or $35.1^\circ$ with the spin axis of this pulsar (see Fig.~\ref{RadioF11}).

Within the framework of the present emission mechanism, the value of ${\hat B}_0$ can be significantly different from that given by the formula for magnetic dipole radiation, in which case equation~(\ref{E27}) and Fig.~\ref{RadioF12} merely determine the required value of ${\hat B}_0 d^2$ as a function of $\alpha$.

In the same way, the values of $\sigma_0$ and $\kappa$ in the caption to Fig.~\ref{RadioF2} together with the period $0.361$ s and the distance $1.26$ kpc yield
\begin{equation}
{\hat B}_0d^2=0.127\,{\hat\kappa}_{\rm th}^{-1/2}\quad{\rm for}\quad J0206-4028.
\label{E28}
\end{equation}
If $d$ is set equal to $1$ and the value of ${\hat B}_0$ is assumed to be that given by the formula for magnetic dipole radiation, i.e.\ $0.879$, then equation~(\ref{E28}) and the red curve in Fig.~\ref{RadioF12} imply that $\alpha$ equals either $0.29^\circ$ or $62.8^\circ$.  According to Fig.~\ref{RadioF11}, the values of $\theta_{P2S}$ corresponding to $\alpha=0.29^\circ$ and $62.8^\circ$ are $0.29^\circ$ and $97.2^\circ$, respectively.

In the case of the broken-power-law example in Fig.~\ref{RadioF3}, the fit parameters $\sigma_0=2.81\times10^{-3}$ and $\kappa=2.81\times10^{-2}$ mJy together with the period $5.16\times10^{-3}$ s and the distance $1.22$ kpc yield
\begin{equation}
{\hat B}_0d^2=1.02\times10^{-3}\,{\hat\kappa}_{\rm th}^{-1/2}\quad{\rm for}\quad J1024-0719.
\label{E29}
\end{equation}
Since ${\hat\kappa}_{\rm th}^{-1/2}\ge1.42$ (see Fig.~\ref{RadioF12}), equation~(\ref{E29}) implies that either $d\ge2.15$ if ${\hat B}_0$ is given by its magnetic-dipole-radiation value $3.13\times10^{-4}$ or ${\hat B}_0\ge1.45\times10^{-3}$ if $d=1$.

In contrast, the fit parameters for the broken-power-law example in Fig.~\ref{RadioF4} together with the period $0.549$ s and the distance $0.4$ kpc yield
\begin{equation}
{\hat B}_0d^2=0.38\,{\hat\kappa}_{\rm th}^{-1/2}\quad{\rm for}\quad J0452-1759,
\label{E30}
\end{equation}
an equation that is satisfied by the magnetic-dipole-radiation value of ${\hat B}_0$ (i.e.\ $1.8$) and $d=1$ if $\alpha=57.7^\circ$ and so $\theta_{P2S}=84.6^\circ$.

The values of the fit parameters $\sigma_0$ and $\kappa$ used for plotting Fig.~\ref{RadioF5} together with the period $0.715$ s and the distance $1.67$ kpc yield
\begin{equation}
{\hat B}_0d^2=25.8\,{\hat\kappa}_{\rm th}^{-1/2}\quad{\rm for}\quad J0332+5434.
\label{E31}
\end{equation}
The constraint ${\hat\kappa}_{\rm th}^{-1/2}\ge1.42$ (see Fig.~\ref{RadioF12}) therefore implies that either $d\ge5.58$, if ${\hat B}_0$ has its magnetic-dipole-radiation value $1.22$, or ${\hat B}_0\ge36.6$, if $d=1$.  

Next, the values of $\sigma_0$ and $\kappa$ in the caption to Fig.~\ref{RadioF6} together with the period $0.105$ s and the distance $7.11$ kpc yield
\begin{equation}
{\hat B}_0d^2=0.207\,{\hat\kappa}_{\rm th}^{-1/2}\quad{\rm for}\quad J1845-0743.
\label{E32}
\end{equation}
In this case, too, the constraint ${\hat\kappa}_{\rm th}^{-1/2}\ge1.42$ (see Fig.~\ref{RadioF12}) implies that either $d\ge1.22$, if ${\hat B}_0$ has its magnetic-dipole-radiation value $0.198$, or ${\hat B}_0\ge0.294$, if $d=1$.  

Unlike the examples in Figs.~\ref{RadioF1}--\ref{RadioF6} for which the value of $j$ in equation~(\ref{E4}) equals $2$, the example in Fig.~\ref{RadioF7} is fitted with a flux density for which $j=0$.  The first member of equation~(\ref{E24}), the values of the fit parameters $\sigma_0$ and $\kappa$ in the caption to Fig.~\ref{RadioF7} and the period $0.307$ s and the distance $5.92$ kpc jointly yield
\begin{equation}
{\hat B}_0d^2=4.03\times10^{-6}\,{\tilde\kappa}_{\rm th}^{-1/2}\quad{\rm for}\quad J1829-1751.
\label{E33}
\end{equation}
The resulting value of ${\tilde\kappa}_{\rm th}^{-1/2}$ for ${\hat B}_0=1.32$ and $d=1$, i.e. $3.28\times10^5$, implies that $\alpha\simeq90^\circ$ in this case (see the curve delineated by the blue dots in Fig.~\ref{RadioF12}).

In the case of the example shown in Fig.~\ref{RadioF8}, too, $j$ equals zero so that the corresponding values of the fit parameters $\sigma_0$ and $\kappa$ together with the period $0.306$ s and the distance $5.03$ kpc yield
\begin{equation}
{\hat B}_0d^2=1.58\times10^{-6}\,{\tilde\kappa}_{\rm th}^{-1/2}\quad{\rm for}\quad J1835-0643.
\label{E34}
\end{equation}
If ${\hat B}_0$ has the value $3.56$ that is obtained from the formula for magnetic dipole radiation, then ${\tilde\kappa}_{\rm th}^{-1/2}=2.25\times10^6$ for $d=1$ and so $\alpha\simeq90^\circ$ according to Fig.~\ref{RadioF12}.

For the double-turn-over spectrum shown in Fig.~\ref{RadioF9}, $j$ equals $2$ and the second member of equation~(\ref{E24}) in conjunction with the values of the fit parameters $\sigma_0$ and $\kappa$, the period $0.277$ s and the distance $0.31$ kpc yield
\begin{equation}
{\hat B}_0d^2=0.198\,{\hat\kappa}_{\rm th}^{-1/2}\quad{\rm for}\quad J1932+1059.
\label{E35}
\end{equation}
If $d$ is set equal to $1$ and the value of ${\hat B}_0$ is assumed to be that given by the formula for magnetic dipole radiation, i.e.\ $0.518$, then equation~(\ref{E35}) and the red curve in Fig.~\ref{RadioF12} imply that $\alpha$ equals $46.5^\circ$.  Moreover, the colatitude along which this pulsar is observed has the value $\theta_{P2S}=60.8^\circ$ according to Fig.~\ref{RadioF11}.

Finally, the values of the fit parameters in the caption to Fig.~\ref{RadioF10}, the period $1.22$ s and the distance $2.3$ kpc yield
\begin{equation}
{\hat B}_0d^2=1.78\times10^2\,{\hat\kappa}_{\rm th}^{-1/2}\quad{\rm for}\quad J0141+6009.
\label{E36}
\end{equation}
Given the constraint ${\hat\kappa}_{\rm th}^{-1/2}\ge1.42$ (see Fig.~\ref{RadioF12}), this implies that either $d\ge19.0$, if ${\hat B}_0$ has its magnetic-dipole-radiation value $0.70$, or ${\hat B}_0\ge2.53\times10^2$, if $d=1$.  

\section{Concluding remarks}
\label{sec:conclusion}

No emission mechanism is as yet identified in the published literature on pulsars whose spectral distribution function can fit the data on all five categories of spectral shapes depicted in Figs.~\ref{RadioF1}--\ref{RadioF10}~\citep[see][and the references therein]{Jankowski, Swainston}.  Curved or gigahertz-peaked spectra are generally thought to reflect the free-free absorption of the pulsar radiation in ionised high-density environments rather than being intrinsic to the emission mechanism~\citep[see][and the references therein]{Rajwade,Kijak}.  As we have seen, however, the spectral distribution function of the caustics that are generated by the superlluminally moving current sheet in the magnetosphere of a non-aligned neutron star single-handedly accounts for all observed features of pulsar spectra (including those that are normally fitted with simple or broken power laws).

A study of the characteristics of the radiation that is generated by this superluminally moving current sheet has already provided an all-encompassing explanation for the salient features of the radiation received from pulsars: its brightness temperature, polarization, spectrum, profile with microstructure and with a phase lag between the radio and gamma-ray peaks~\citep{Ardavan2021, Ardavan2022a} and the discrepancy between the energetic requirements of its radio and gamma-ray components~\citep{Ardavan2023}.  Fits to the exceptionally broad gamma-ray spectra of the Crab, Vela and Geminga pulsars, for example, are provided by the spectral energy distribution of this radiation over the entire range of photon energies so far detected from them~\citep{Ardavan2023Crab,ArdavanCVG}. 

Detailed analyses of the structure of the magnetospheric current sheet and the coherent emission mechanism by which this sheet creates the caustics underlying the present spectral distribution function can be found in~\citet{Ardavan2021}.  A heuristic account of the mathematical results of those analyses in more transparent physical terms is presented in~\citet{Ardavan2022a} and~\citet[][Section~2]{Ardavan2023Crab}.

Finally, the following cautionary remark concerning a common misconception is in order: it is often presumed that the plasma equations used in the numerical simulations of the magnetospheric structure of an oblique rotator should, at the same time, predict any radiation that the resulting structure would be capable of emitting~\citep[see, e.g.][]{SpitkovskyA:Oblique,Contopoulos:2012}.  This presumption stems from disregarding the role of boundary conditions in the solution of Maxwell’s equations.  As we have already pointed out, the far-field boundary conditions with which the structure of the pulsar magnetosphere is computed are radically different from the corresponding boundary conditions with which the retarded solution of these equations (i.e. the solution describing the radiation from the charges and currents in the pulsar magnetosphere) is derived (see Section~3 and the last paragraph in Section~6 of~\citealt{Ardavan2021}).

\section*{Acknowledgements}
 I thank N.\ A.\ Swainston for helpful correspondence.
\addcontentsline{toc}{section}{Acknowledgements}

\section*{Data availability}

The data used in this paper are available in the public domain.

\bibliographystyle{mnras}
\bibliography{RadioSpec.bib}







\bsp	
\label{lastpage}
\end{document}